\documentclass[sigconf,10pt]{acmart}
\usepackage{booktabs} 
 \usepackage{enumitem}
\usepackage{balance}
\usepackage[linesnumbered,ruled,vlined]{algorithm2e}
\usepackage{xcolor}
\usepackage{listings}
\usepackage{cleveref}
\usepackage{graphicx}
\usepackage{multirow}

\setcopyright{none} 
\renewcommand\footnotetextcopyrightpermission[1]{} 
\SetKwInput{KwInput}{Input}                
\SetKwInput{KwOutput}{Output}              

\SetCommentSty{commfont}

\hypersetup{draft}
\begin{document}
\title{AI Powered Compiler Techniques for DL Code Optimization}


\author{Sanket Tavarageri}
\affiliation{
  \institution{Intel Labs}
  \country{India}
}
\email{sanket.tavarageri@intel.com}

\author{Gagandeep Goyal}
\affiliation{
  \institution{IIT Hyderabad}
  \country{India}
    }
\email{cs19mtech01003@iith.ac.in}

\author{Sasikanth Avancha}
\affiliation{
  \institution{Intel Labs}
  \country{India}
}
\email{sasikanth.avancha@intel.com}

\author{Bharat Kaul}
\affiliation{
  \institution{Intel Labs}
  \country{India}
}
\email{bharat.kaul@intel.com}

\author{Ramakrishna Upadrasta}
\affiliation{
  \institution{IIT Hyderabad}
  \country{India}
    }
\email{ramakrishna@iith.ac.in}


\begin{abstract}
Creating high performance implementations of deep learning primitives on CPUs is a challenging task. Multiple considerations including multi-level cache hierarchy, and wide SIMD
units of CPU platforms influence the choice of program transformations to apply for performance optimization. 
In this paper, we present machine learning powered compiler techniques to optimize loop nests.
We take a two-pronged approach to code optimization: We first apply high level optimizations to optimize the code to take optimal advantage of the cache memories.
Then, we perform low level, target-specific optimizations to effectively vectorize the code to run well on the SIMD units of the machine. For high level optimizations, we use polyhedral compilation techniques and deep learning approaches. For low level
optimization, we use a target specific code generator that generates code using vector intrinsics and Reinforcement Learning (RL) techniques to find the optimal parameters for
the code generator. We perform experimental evaluation of the developed techniques
on various matrix multiplications that occur in popular deep learning workloads.
The experimental results show that the compiler techniques presented in the paper
achieve 7.6X and 8.2X speed-ups over a baseline for sequential and parallel runs respectively.
\end{abstract}

%
%

%
%

\settopmatter{printfolios=true}
\maketitle

\pagestyle{plain} 

\section{Introduction}
\label{sec:intro}

Deep learning (DL) has become pervasive in various domains of computing. Image recognition \cite{krizhevsky2012imagenet,he2016deep}, language modeling \cite{devlin2018bert}, language translation \cite{wu2016google}, speech recognition \cite{hinton2012deep} make extensive use of deep neural networks (DNNs). 
Deep Learning inference is an important workload across applications, such as object classification \& recognition, text and speech translation etc.
A 2018 Mc Kinsey study \cite{mckinseystudyinferencehardware} pointed out that in datacenters, 75\% of the inference tasks are run on CPUs. Optimizing DL workloads on CPUs is a challenging proposition because of architectural complexities of CPU platforms. Multi-level cache hierarchies, TLBs (Translation Look-aside Buffers), hardware data prefetchers, SIMD (a.k.a vector) units present particular challenges in writing high performance code. Therefore, the current state-of-practice is to use expert-coded high performance libraries such as Intel oneDNN \cite{intelmkldnn} in deep learning frameworks such as TensorFlow and PyTorch to achieve good performance. However, being reliant on libraries for performance is not scalable. First, it would increase the time-to-market: from the time a new DL operator is invented to its being supported in a library could take a considerable amount of time. Second, even expert programmers must invest significant amount of effort to tune the implementations on the target platforms. Therefore, an attractive alternative solution is to develop compilation techniques that  automate code optimization and achieve similar performance levels as expert-coded libraries.

In this paper, we develop a systematic approach to automatic code optimization. We categorize the program optimizations into two phases: high level and low level optimizations. High level optimizations perform loop optimizations such as loop reordering and tiling to derive a loop structure that utilizes the cache hierarchy of the computer system to the fullest extent possible. Low level optimizations generate 
vector code using the target machine's \emph{intrinsics}; reinforcement learning methodology guides the derivation of high performance vector code. For high-level and low-level optimizations we leverage artificial intelligence (A.I.) techniques.
We evaluate our automated compiler system on GEMMs which lie at the heart of deep learning \cite{gemmdl}.
The results indicate our compiler workflow delivers competitive performance compared to
 Intel oneDNN library and significantly higher performance compared to a 
 state-of-the-art DL compiler, viz., AutoTVM \cite{chen2018learning}. 

The contributions of the paper are as follows.
\begin{itemize}
 \item A systematic approach to program optimization with clear demarcation of high-level and low-level optimizations that map well to the hardware architectures.
 \item Low-level optimizations that generate reinforcement learning (RL) guided vector \emph{intrinsics} based code.
 \item Development of A.I. techniques for high-level and low-level optimizations.
 \item Experimental evaluation on various GEMM sizes that occur in DL workloads.
\end{itemize}

The rest of the paper is organized as follows. Section \ref{section:workflow} introduces the overall compilation workflow. Section \ref{section:hlo} describes the polyhedral compilation techniques we use for high level optimizations.
The low-level optimizations involving the target platform specific code generator and reinforcement learning are developed in Section \ref{section:llo}. 
The experimental evaluation conducted is detailed in Section \ref{section:experiments}.
Related work is discussed in Section \ref{sec:related}. The conclusion and implication
of the presented work are presented in Section \ref{sec:conclusion}.

\newcommand{\fixme}[1]{\textcolor{red}{\textbf{URK FIXME:}#1}}

\newcommand{\gagancomment}[1]{\textcolor{blue}{\textbf{GAGAN COMMENT:}#1}}

\begin{figure*}[t!]
\centering
\centering
\includegraphics[scale=0.7]{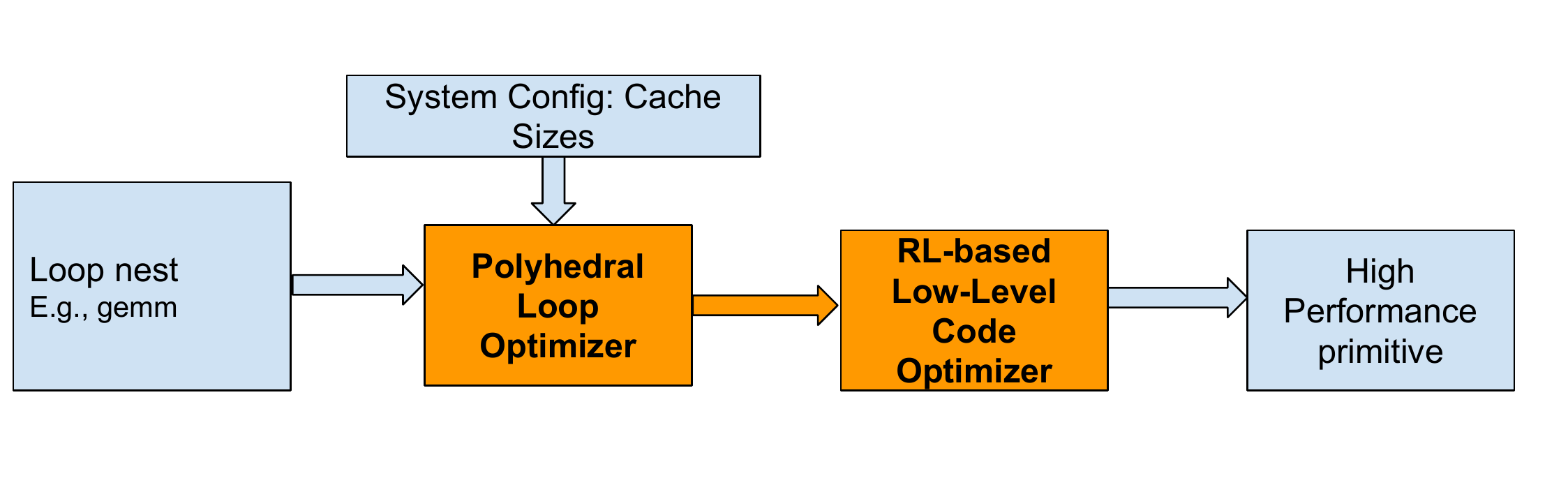}
\caption{The compiler optimizations workflow}
\label{fig:polydl_workflow}
\end{figure*}

\section{The Compiler Optimization Workflow}
\label{section:workflow}

We first describe the overall compiler workflow.
We input the loop nests such as GEMMs to the compiler.
The high level optimizer first optimizes the loop structure and then passes on
the code to the low level optimer.
Figure \ref{fig:polydl_workflow} shows the workflow of the optimization 
process.

The high level optimizer uses polyhedral compilation techniques for optimization of the loop structure to 
take advantage of the multi-level caches of the CPU platform.
The loop reordering and tiling transformations are applied and
the best loop order and tile sizes are determined by the high level optimizer.
It will enhance data locality -- both the spatial and temporal locality such that
the data used by the input program is reused out of the caches closest to the processor
as much as possible. 
Section \ref{section:hlo} details the polyhedral compilation techniques we apply
for loop optimization.

The high level optimer then hands over the optimized code to the low-level optimizer.
The low level optimizer derives a vectorization strategy for effective use of 
the SIMD vector units. We employ  a target specific low level optimization approach
wherein the inner loops of the loop nest are vectorized using the target-specific
vector intrinsics. Further, 
we use a Reinforcement Learning (RL) based approach to select the best vectorization plan among the myriad choices available.

Jouppi et al \cite{jouppi2017datacenter} show that 95\% of the deep learning
inference workloads (MLPs, CNNs, and LSTMs) can be formulated in terms of matrix-multiplication. Matrix multiplication is at the heart of deep learning \cite{gemmdl}.
Because of these reasons in this work, we build the low level optimizer to optimize
the matrix-multiplications that occur in the inner most loops of convolutions and matrix-multiplications themselves (matrix-multiplication can be recursively defined where the other loops of the code are the tiled loops and the inner loops are also functionally equivalent to matrix-multiplication).
In Section \ref{section:llo}, we describe the design and the implementation of the 
low level optimizer for the inner loops of the loop nest focused on matrix multiplication.

\section{High-level Polyhedral Loop Optimizations}
\label{section:hlo}
We use the polyhedral model \cite{feautrier1996automatic}, which is an advanced mathematical framework to reason about dependences and loop transformations, to develop our data reuse algorithm. 

\subsection{Preliminaries}
We use the Integer Set Library (ISL) \cite{verdoolaege2010isl} for performing polyhedral operations in this work and we use the same notation as used in ISL to elucidate the concepts and the algorithm.
The matrix multiplication code shown in Figure \ref{matmulcode} will be used to illustrate the workings of the data reuse analysis.

\begin{figure}
\begin{lstlisting}[language=C,basicstyle=\small,frame=bottomline]
for (i = 0; i < M; i++) {
    for (j = 0; j < N; j++) {
	    for (k = 0; k < K; k++) {
    	    C[i][j] += A[i][k] * B[k][j];
        }
    }
}
\end{lstlisting}
\caption{Matrix multiplication code}
\label{matmulcode}
\end{figure}

\paragraph{Sets}
A set is a tuple of variables $x_i$s along with a collection of constraints $c_k$s defined on the tuple variables. 
$ s = \{ [x_1, \dots , x_n] : c_1 \land \dots c_m \} $

The iteration spaces of loop nests are represented as sets. The iteration space of the loop in Figure \ref{matmulcode} is defined as the following set.
$ I = \{ S[i, j, k] : 0 <= i < M \land 0 <= j < N \land 0 <= k < K \} $

\paragraph{Relations}
A relation is a mapping from input tuple variables $x_i$s to output tuple variables $y_j$s.
In addition, a set of constraints $c_k$s can be defined for a relation that will place constraints on the input/output tuple variables. 
$ r = \{ [x_1, \dots, x_n] \mapsto [y_1, \dots, y_m] :  c_1, \dots, c_p \}$

The read and write access functions of a loop nest can be modeled with relations. The read relations in the Figure \ref{matmulcode} code are shown below: 
$r_1  = \{ S[i, j, k] \mapsto  C[i, j] \}$,
$r_2  = \{ S[i, j, k] \mapsto  A[i, k] \}$,
$r_3  = \{ S[i, j, k] \mapsto  B[k, j] \}$.
The sole write relation in the loop is: $w_1 = S[i, j, k] \mapsto C[i, j]$.
The domain of a relation $r$ is denoted by {\textsf dom} $r$.

\paragraph{Apply operation} When a relation $r$ is applied on a set $s$, the domain of $r$ will be intersected with $s$ and the resulting range will be a new set $s'$. The set $s'$ is said to be the result of the apply operation. The operation is mathematically defined as:
$ ( \vec{y} \in s') \Longleftrightarrow (\exists \vec{x} ~~\text{s.t}~~ (\vec{x} \in s \land \vec{x} \mapsto \vec{y}) \in r )$

The data footprint of the loop can be computed by \emph{applying} read and write \emph{relations} on the iteration space \emph{set}:
$ r_1(I) \cup r_2(I) \cup r_3(I) \cup w_1(I) $

\paragraph{Lexicographic operations} The lexicographical operations can be applied on
 sets. $s_1 << s_2$ outputs all the elements of $s_1$ that are  lexicographically 
strictly smaller than all the elements of $s_2$, while $s_1 <<= s_2$ gets us the elements of $s_1$ that are lexicographically smaller than or equal to the elements of $s_2$.
The lexicographically smallest element of a set $s$ is queried using {\textsf lexmin} $s$.
Similarly, the lexicographically largest element is obtained using {\textsf lexmax} $s$.

\paragraph{Set difference.} The set difference
between set $s_1$ and $s_2$ is denoted by $s_1 - s_2$, i.e., the resulting set will have
elements of $s_1$ that do not appear in $s_2$.

\paragraph{Polyhedral dependences.}
The exact data dependences in loop nests can be computed in the polyhedral model and are expressed as maps from source iterations to target iterations involved in the dependence. For cache data reuse analysis developed in \S\ref{sec:wscompute}, we consider  four kinds of dependences -- Read-After-Read (RAR), Read-After-Write (RAW, a.k.a \emph{flow}), Write-After-Read (WAR, a.k.a \emph{anti}), and Write-After-Write (WAW). 
  The data dependencies of the matrix multiplication code in Figure \ref{matmulcode} are shown below.
  
\begin{align*}
d_1 = & \{ S[i, j, k] \mapsto S[i', j', k'] : i' = i \land j' = j \land k < k' < K \} \\
d_2 = & \{ S[i, j, k] \mapsto S[i', j', k'] : i' = i \land k' = k \land j < j' < N \} \\
d_3 = & \{ S[i, j, k] \mapsto S[i', j', k'] : j' = j \land k' = k \land  i < i' < M \} \\
\end{align*}

The dependence $d_2$ is induced by array reference A[i][k]. An element of array A, say A[0][0] which is accessed in \emph{source} iteration $[i=0,j=0,k=0]$ gets reused in \emph{target} iterations $[i'=0,j'>0,k'=0]$. The source to target iteration relationships such as this are expressed in a parametric fashion as the relation $d_2$. 

\subsection{Loop transformations}
\label{sec:wscompute}
We create a number of code variants by applying loop reordering and tiling transformations and using the PolyDL techniques \cite{tavarageri2021polydl} select the top code variants.

\paragraph{Working set size computation}
We perform cache data reuse analysis to characterize a loop-nest's behavior with respect to a given cache hierarchy. The analysis computes the various existing data reuses of a program and then for the input cache hierarchy determines which data reuses are exploitable at various levels of cache.
Each data dependence in a loop is also an instance of data reuse -- the source and target iterations involved in the dependence touch the same data element and therefore, the data is reused. For a data dependence and hence data reuse to be realizable in a given level of cache, all the data elements accessed between the source and target iterations of the dependence -- the \emph{working set} -- have to be retained in the cache so that when the execution reaches the target iteration, the data element(s) used in the source iteration will still be present in the cache. 

We illustrate the computation of the working set sizes using the running example in Figure \ref{matmulcode}. Let us examine the following dependence carried by the $j$ loop arising because of the array reference $A[i][k]$:
$d_2 =  \{ S[i, j, k] \mapsto S[i', j', k'] : i' = i \land k' = k \land j < j' < N \}$.
Of all the source iterations, the first/lexicographically minimum iteration is: 
$\mathcal{I}_{source} = \{ S[i = 0, j = 0, k = 0] \}$
Its target iterations are: 
$\{ S[i = 0, j, k = 0] :  0 < j < N \}$.
Among the target iterations, the first one is: 
$I_{min\_tar} = \{ S[i = 0, j = 1, k = 0]  \}$ 
and the last one is: 
$I_{max\_tar} = \{ S_3[i = 0, j = N-1, k = 0]  \}$

The number of data elements of the three arrays -- A, B, C accessed between $\mathcal{I}_{source}$ and $I_{min\_tar}$ is derived by \emph{applying} the read and write relations on the intervening iteration set and it is: $$WS_{min} = 2K + 3$$

The $K$ elements of array A -- $A[0][0, 1, \dots, K-1]$, the $K+1$ elements of array B --
$B[0, 1, \dots, K-1][0]$ and $B[0][1]$, and finally $2$ elements of array C -- 
$C[0][0], C[0][1]$ accessed between the source iteration $S[i = 0, j = 0, k = 0]$
and the target iteration $I_{min\_tar} = S[i = 0, j = 1, k = 0]$ lead to the $WS_{min}$ size of $2K + 3$.

The maximum working set size -- the number of data elements touched between $\mathcal{I}_{source}$ and $I_{max\_tar}$ is:

$$ WS_{max} = N\times K + N +1 $$
The $WS_{max}$ size is arrived at by counting the number of array elements
accessed between the source iteration - $S[i = 0, j = 0, k = 0]$ and the target iteration - 
$I_{max\_tar} = \{ S_3[i = 0, j = N-1, k = 0]  \}$.
As far as array A is concerned, $K$ elements of it -- $A[0][0, 1, \dots, K-1]$ are read.
Array B's elements -- $B[0, 1, \dots, K-1][0, 1, \dots, N-2]$ plus $B[0][N-1]$ are read which total $K \times (N-1) + 1$.
$N$ elements of array C are read and written -- $C[0][0, 1, \dots, N-1]$. Therefore, a total of $N\times K + N +1$ are read and written.

We have built a code generator to emit a number of program variants.
The code generator creates the loop variants by applying tiling and loop interchange
program transformations. The tile sizes are varied as well.
The working set size computation analysis is performed on each program version generated.
Among the many variants generated, the ranking algorithm described below picks
the top $k$ best performing versions, where $k$ is a parameter.

\paragraph{DNN-based code ranking algorithm.}
We assume fully associative, and exclusive caches. 
If the working set size corresponding to a data reuse in the program is smaller
than the cache size then the data reuse is exploitable in the cache.
The ranking system considers caches at different levels (typically L1, L2, and L3)
and for each data reuse, determines at what level of cache hierarchy is the data reuse
realizable.
We now describe the algorithm to determine the cumulative
working set sizes at each level of cache. The inputs to the algorithm are the 
working set sizes computed for a loop nest, and the cache sizes of the target system.
The algorithm determines the fastest level of cache where the working set size corresponding to each data reuse fits
and adds it to that cache's working set size. 
If a working set does not fit in any cache, then the data reuse happens
out of the main memory. Consequently, the memory's working set size is updated.

We use a deep neural network (DNNs) for 
ranking of code variants. 
For the purposes of training the DNN model, 
we collect the performance data of code variants generated and
compute their working set sizes at different levels of the memory hierarchy.
We train the DNN model to perform relative ordering of \emph{two} code
variants.
We then use a \emph{tournament} based ranking system to assign ranks
to the different code versions created -- 
we play each code variant against every other code variant.
For each variant, we record the number of wins it has accumulated.
We then rank the variants based on the number of wins -- 
the higher the number of wins, the higher the rank.

\begin{figure}
\centering
\includegraphics[scale=0.6]{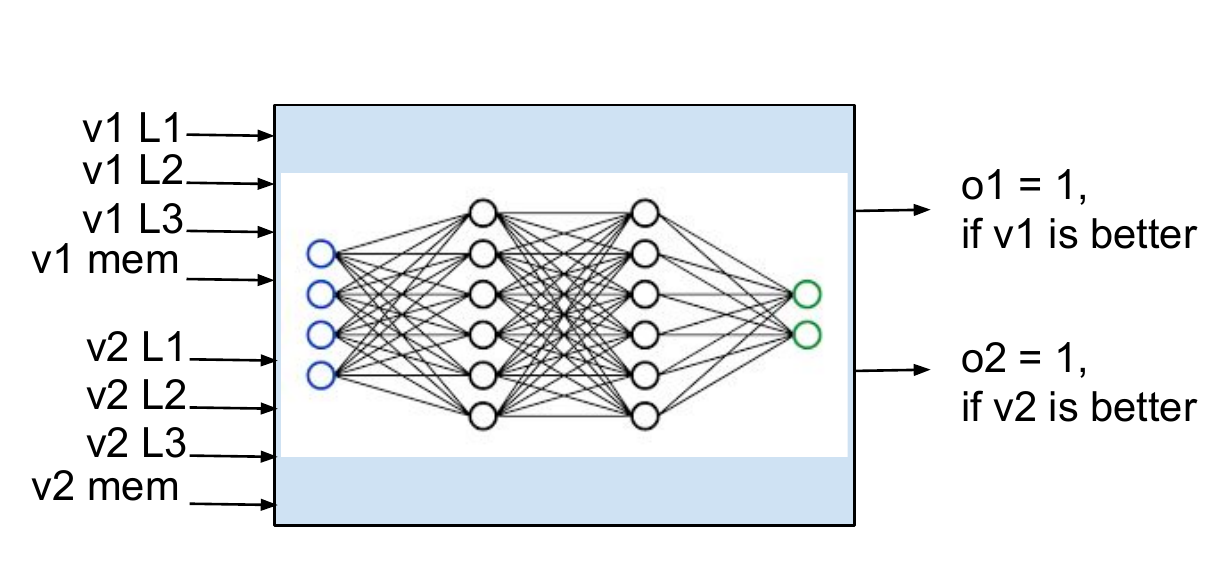}
\caption{The DNN architecture for ranking of code variants. The DNN shown is a representative figure.}
\label{fig:polydnn}
\end{figure}

We use a four layer feed forward neural network architecture shown in Figure \ref{fig:polydnn}.
We normalize the compiler generated working set sizes using min-max scaling: 
$\frac{x - x_{min}}{x_{max} - x_{min}}$. 
Each value is subtracted with the minimum value in that feature column and divided by the feature range.
The output layer consists of two neurons and we use the \emph{softmax}
function for the output layer.
The values of the two output neurons, because of the use of the softmax function,
sum to 1. 
If the output value is above a threshold - $\theta$, we consider it a 1, otherwise a 0.
If the first neuron fires a 1, then the first variant is considered the winner.
If the second neuron fires a 1, then the second variant is considered the winner.
If both of them are zero because none of them are above the threshold, then
it is a draw between the two variants. In this work, we set the threshold $\theta$ to 0.7.
We experimented with deeper models as well. However,  depth beyond 
four did not have any discernible effect on accuracy.

\section{Low-level Target Specific Inner Loop Optimizations}
\label{section:llo}
The high level optimizations as described in \S \ref{section:hlo} are first applied to the input code and then the inner loops 
are handed over the low level optimizer.
The low level optimizer focuses on vectorization and assumes that the data 
used by the inner loops is resident in L1 cache.
The inner loops are analyzed to find out which loops are parallel and hence,
vectorizable. 
The different vectorizable loops present us multiple choices for vectorization.
Further, unroll-and-jam (the loops are unrolled and the unrolled statements are combined
in the inner-most loop) can present various data reuse opportunities.
Thus, the various unroll factors for the loops
give rise to multiple ways of vectorizing the loops and we have to select a scheme 
that leads to the highest performance. To help select the best vectorization parameters,
namely, the unroll factors for the loops, we use Reinforcement Learning (RL).

\begin{figure}
\begin{lstlisting}[language=C,basicstyle=\small,frame=bottomline]
for (i = 0; i < M; i++) {
 for (j = 0; j < N; j+=16) {
  for (k = 0; k < K; k++) {
   C[i][j] += A[i][k] * B[k][j];
   C[i][j+1] += A[i][k] * B[k][j+1];
   C[i][j+2] += A[i][k] * B[k][j+2];
   ...
   C[i][j+15] += A[i][k] * B[k][j+15];
  }
 }
}
\end{lstlisting}
\caption{Unrolled GEMM code for unroll factors 1, 16, 1}
\label{matmulunrolled}
\end{figure}

\begin{figure}
\begin{lstlisting}[language=C,basicstyle=\small,frame=bottomline]
M_full = (M / 1) * 1 ;
N_full = (N / 16) * 16 ;
K_full = (K / 1) * 1 ;
for (i = 0; i < M_full; i += 1) {
 for (j = 0; j < N_full; j += 16) {
  vecC = _mm512_load_ps(&C[i*CStride+j]);
  for (k = 0; k < K_full; k += 1) {
   vecA =_mm512_set1_ps(A[i*AStride + k]);
   vecB =_mm512_load_ps(&B[k*BStride+j]);
   vecC =_mm512_fmadd_ps(vecA,vecB,vecC);
  }
  _mm512_store_ps(&C[i*CStride + j], vecC);
 }
}
// The residue code for non-full M, N, K 
// values omitted for brevity.
\end{lstlisting}
\caption{Auto-generated GEMM code using AVX-512 intrinsics for unroll factors 1, 16, 1}
\label{matmulunrolledAVX1}
\end{figure}

\begin{figure}
\begin{lstlisting}[language=C,basicstyle=\small,frame=bottomline]
M_full=(M / 2) * 2 ;
N_full=(N / 16) * 16 ;
K_full=(K / 2) * 2 ;
for (i=0; i < M_full; i+=2) {
 for (j=0; j < N_full; j+=16) {
  vecC=_mm512_load_ps(&C[i*CStride+j]);
  vecC1=_mm512_load_ps(&C[(i+1)*CStride+j]);
  for (k=0; k < K_full; k+=2) {
  vecA=_mm512_set1_ps(A[i*AStride+k]);
  vecB=_mm512_load_ps(&B[k*BStride+j]);
  vecB1=_mm512_load_ps(&B[(k+1)*BStride+j]);
  vecA1=_mm512_set1_ps(A[i*AStride+(k+1)]);
  vecA2=_mm512_set1_ps(A[(i+1)*AStride+k]);
  vecA3=_mm512_set1_ps(A[(i+1)*AStride+(k+1)]);
  vecC=_mm512_fmadd_ps(vecA,vecB,vecC);
  vecC1=_mm512_fmadd_ps(vecA2,vecB,vecC1);
  vecC=_mm512_fmadd_ps(vecA1,vecB1,vecC);
  vecC1=_mm512_fmadd_ps(vecA3,vecB1,vecC1);
  }
 _mm512_store_ps(&C[i*CStride+j], vecC);
 _mm512_store_ps(&C[(i+1)*CStride+j], vecC1);
 }
}
// The residue code for non-full M, N, K 
// values omitted for brevity.
\end{lstlisting}
\caption{Auto-generated GEMM code using AVX-512 intrinsics for unroll factors 2, 16, 2}
\label{matmulunrolledAVX2}
\end{figure}

\subsection{Vector intrinsic based code generation}
\label{sec:veccodegen}
We illustrate the workings of the vectorization scheme and the use of RL on 
GEMM inner loops. Figure \ref{matmulunrolled} shows the matrix multiplication
code where the \emph{j} loop is unrolled by a factor of 16 and the statements
are moved to the inner most loop (unroll-and-jam).
 Because the \emph{j}
loop is parallel, it is vectorizable. We have built a code generator
that generates the vectorized code using the \emph{vector intrinsics} of the 
target CPU platform. Figure \ref{matmulunrolledAVX1} shows
the generated code using AVX-512 intrinsics to run on vector units
that can work on 512 bits of data simultaneously.
The datatype of the variables in the shown code is 32 bit floating point numbers.
Therefore, we can perform arithmetic operations on 16 floating point numbers
($16 \times 32 = 512$) at the same time.
In Figure \ref{matmulunrolled}, we observe that the same array element -- ``A[i][k]''
is used in all 16 arithmetic operations. Therefore, it is broadcast to all elements
of the vector register using the \textsf{\_mm512\_set1\_ps} vector intrinsic.
The 16 elements of the C array -- ``C[i][j]'' through ``C[i][j+15]'' are loaded using the
\textsf{\_mm512\_load\_ps} vector intrinsic.
Since the loaded C elements are reused in all of the inner-most \emph{k} loop,
the loading is hoisted out of the \emph{k} loop.
In a similar fashion, the 16 elements of the B array -- ``B[k][j]'' through ``B[k][j+15]'' are loaded using the
\textsf{\_mm512\_load\_ps} vector intrinsic.
The 16 addition and multiplication operations are performed using 
the fused-multiply-add operation through the intrinsic \textsf{\_mm512\_fmadd\_ps}.
After the C vector is accumulated into in the \emph{k} loop, 
the results are stored back into the C array using \textsf{\_mm512\_store\_ps}
outside of the \emph{k} loop.

In the GEMM code, all three loops -- \emph{i},\emph{j}, and \emph{k} loops carry data reuse. Consequently, when we unroll any of the loops, that leads to reuse of the data
of one of the three arrays -- A, B, and C.
For example, in Figure \ref{matmulunrolledAVX2}, the \emph{i} loops is unrolled by a factor of 2 and it leads to reuse of the array B: The array access expression for B is
``B[k][j]'' and it is free of the \emph{i} loop variable and therefore for all values of
\emph{i}, the same B array element would be used.
In the code we observe that \textsf{vecB} is used in two fused-multiply-add operations --
while computing \textsf{vecC} and \textsf{vecC1}.
When we unroll the inner-most \emph{k} loop, that helps us schedule 
the \textsf{load} and \textsf{fma} operations further apart thereby increasing the chances of 
the data being loaded in the vector registers when the execution reaches the \textsf{fma} operations.

Thus, by unrolling various loops we can increase the data reuse in vector registers,
and schedule the load operations in such a way that memory latency is tolerated well.
However, because the number of vector registers is limited, a certain choice of an unroll factor for a loop
will impose constraints on the unroll factors of other loops. The interplay between
the amount of data reuse, scheduling, and the impact of data reuse of different arrays could be complex.
We use reinforcement learning which in turn uses a neural network to determine the best
unroll factors for the loops.

\subsection{Reinforcement Learning}
\label{sec:RL}

\begin{figure}
\centering
\includegraphics[scale=0.6]{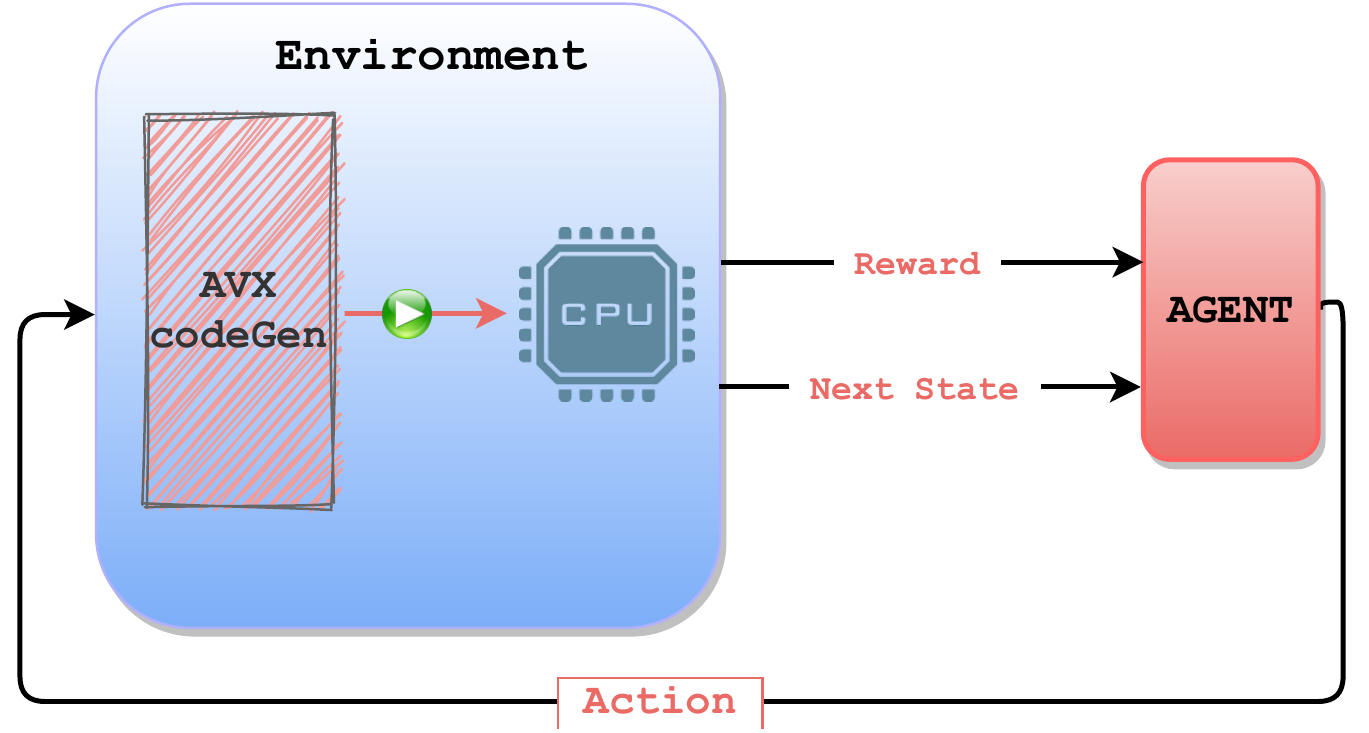}
\caption{Reinforcement Learning for target specific low level optimization.}
\label{fig:llorl}
\end{figure}

The choices of unroll factors for the inner loops constitute the state space
for reinforcement learning (RL). The agent will suggest whether to increase
unroll factors or to decrease them. The increment/decrement of the unroll factors
form the actions. The target specific code generator will carry out the actions suggested
by the agent by generating code with the new unroll factors and the code is run
on the target platform. The actions will either lead to a higher performance  or a lower performance vis-a-vis the performance of the prior state. If the action leads to a higher performance, 
we will encode it as a positive reward -- the relative performance increase.
If the action causes the performance to degrade, it is denoted as a negative reward -- the relative performance decrease. Figure \ref{fig:llorl} shows the RL set-up.

The agent will use a neural network to suggest next actions to undertake. While the state space exploration is being conducted, we will have two phases -- \emph{exploration}, and \emph{exploitation}.
In the exploration phase, the agent will recommend random actions and the reward
obtained will be used to continually train the neural network to predict actions that will lead to larger positive rewards and thus higher performance states.
In the exploitation phase, the agent will query the neural network for the best actions -- actions that will lead to the biggest rewards.
The transitions between the  exploration and exploitation phases are controlled by the \emph{exploration decay rate}. We set it in such a way that at initial stages exploration is selected more often,
and later exploitation is chosen more.

We train a neural network to encode the policy for RL -- whether to increment the unroll factors or two decrement them. The neural network comprises of six intermediate layers  -- two blocks of Dense, Batch Normalization and Dropout layers. For Dense layers we use \emph{Relu} as the activation function. We set the drop-out rate of 0.25 for Dropout layers to avoid overfitting. The output layer is a dense layer which has as many neurons as the number of actions. For matrix multiplication, there are 7 actions possible:
2 actions for each unroll factor (whether to increment or to decrement) and a special state to indicate no further action is necessary.

\begin{figure*}
\centering
\includegraphics[scale=0.8]{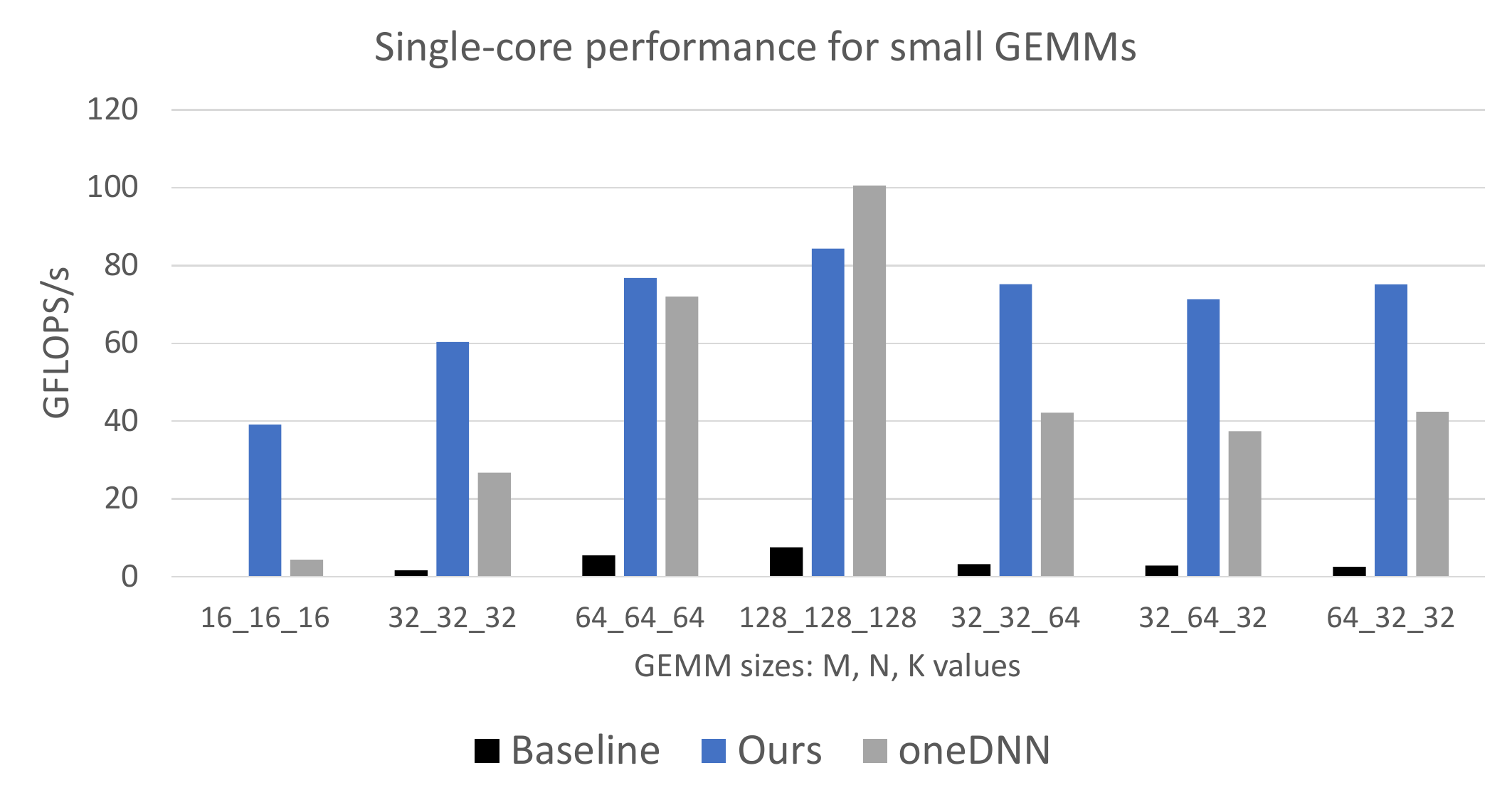}
\caption{Performance of our inner loop optimization using RL on matrices that fit in private L1/L2 caches.}
\label{fig:parallelperf}
\end{figure*}

\section{Experimental Evaluation}
\label{section:experiments}

We evaluate the performance of our compiler framework on the GEMM operation for a range of matrix sizes.
The GEMM operation is, C = A.B where A, B, and C are matrices. Matrix C is of size $M \times N$, A of size $M \times K$ and B of size $K \times N$.
We compare our compiler optimizations against three other systems: 1) The matrix multiplication code shown in Figure \ref{matmulcode} optimized to the highest levels using the Intel C compiler -- icc version 19.1.3.304 with the optimization flag -O3. For parallel runs, we parallelize the outermost -- `i' loop with OpenMP pragmas. We consider the resulting performance to be the \textbf{baseline}. 2) The latest version of Intel oneDNN library version v2.2. The oneDNN library is an expert coded library for inter alia, GEMMs. 3) The latest release of TVM \cite{chen2018tvm} v0.7.0. We use the optimization guide published on the TVM website for GEMMs \cite{autotvmgemmcpu} to obtain its performance. Additionally, we tune the performance of TVM by exploring a number of tile sizes and report the best performance among the different tile sizes explored. The experiments are run on Intel(R) Xeon(R) Platinum 8280 (Cascade Lake) servers running at 
the frequency of 2.70 GHz.
A single socket processor has 28 cores, 32KB private L1 cache, 1MB private L2 cache, and
39MB shared L3 cache.

\subsection{Evaluation of Low-level optimizations}
We first assess the efficaciousness of the low-level optimization scheme we described in Section \ref{section:llo}. To do so, we select matrix sizes such that all the matrices will fit in in the private L1/L2 cache of the processor and thus the code does not require any loop transformations to enhance data locality. 

We show the performances achieved on a single core of the Xeon processor in Figure \ref{fig:parallelperf} by the ICC compiler -- \emph{baseline}, our toolchain, and the oneDNN library. We run each code a 100 times and report the average performance observed. The baseline performance ranges from 0.18 GFLOPS/s (for M = 16, N = 16, and K = 16) to 7.53 GFLOPS/s (for M = 128, N = 128, and K = 128). The peak performance of a core of the said Xeon system is \textasciitilde 118 GFLOPS/s. Thus,
without any further optimizations such as the ones we have described in the paper, we observe that we obtain very low performance.
The oneDNN performance for all problem sizes save for M = N = K = 128 is lower than ours. 
Our toolchain reaches a sizable percentage of the machine's peak performance. 
The lowest performance gotten is 39.10 GFLOPS/s for M = N = K = 16, and the highest 
performance reached is 84.34 GFLOPS/s for M = N = K = 128. As the problem size increases, 
the work to be performed increases, and due to Instruction Level Parallelism (ILP) and because the SIMD units can be kept more busy, the performance goes up.
The experiments show that our target-specific code generation and reinforcement learning scheme is extremely effective in vectorizing the code well.

\begin{table}[t]
\caption{\label{tab:GEMMsizes} GEMM sizes for performing matrix multiplication. Source: Qin et al \cite{qin2020sigma}.}
\small
\begin{tabular}{|c|c|r|r|r|}
\hline
Workload & Application & M & N & K  \\ \hline
\multirow{4}{*}{GNMT} & \multirow{4}{*}{Machine translation} & 128 & 2048 & 4096 \\ \cline{3-5}
 &  & 320 & 3072 &  4096 \\ \cline{3-5}
  &  & 1632 & 36548 &  1024 \\ \cline{3-5}
   &  & 2048 & 4096 &  32 \\ \hline
 \multirow{2}{*}{DeepBench} & \multirow{2}{*}{General workload} & 1024 & 16 & 500000 \\ \cline{3-5}
 & & 35 & 8457 & 2560 \\ \hline
 \multirow{2}{*}{Transformer} & \multirow{2}{*}{Language Understanding} & 31999 & 1024 & 84 \\ \cline{3-5}
 & & 84 & 1024 & 4096 \\ \hline
 \multirow{2}{*}{NCF} & \multirow{2}{*}{Collaborative Filtering} & 2048 & 1 & 128 \\ \cline{3-5}
  & & 256 & 256 & 2048 \\ \hline 
\end{tabular}
\end{table}

\begin{figure*}
\centering
\includegraphics[scale=0.8]{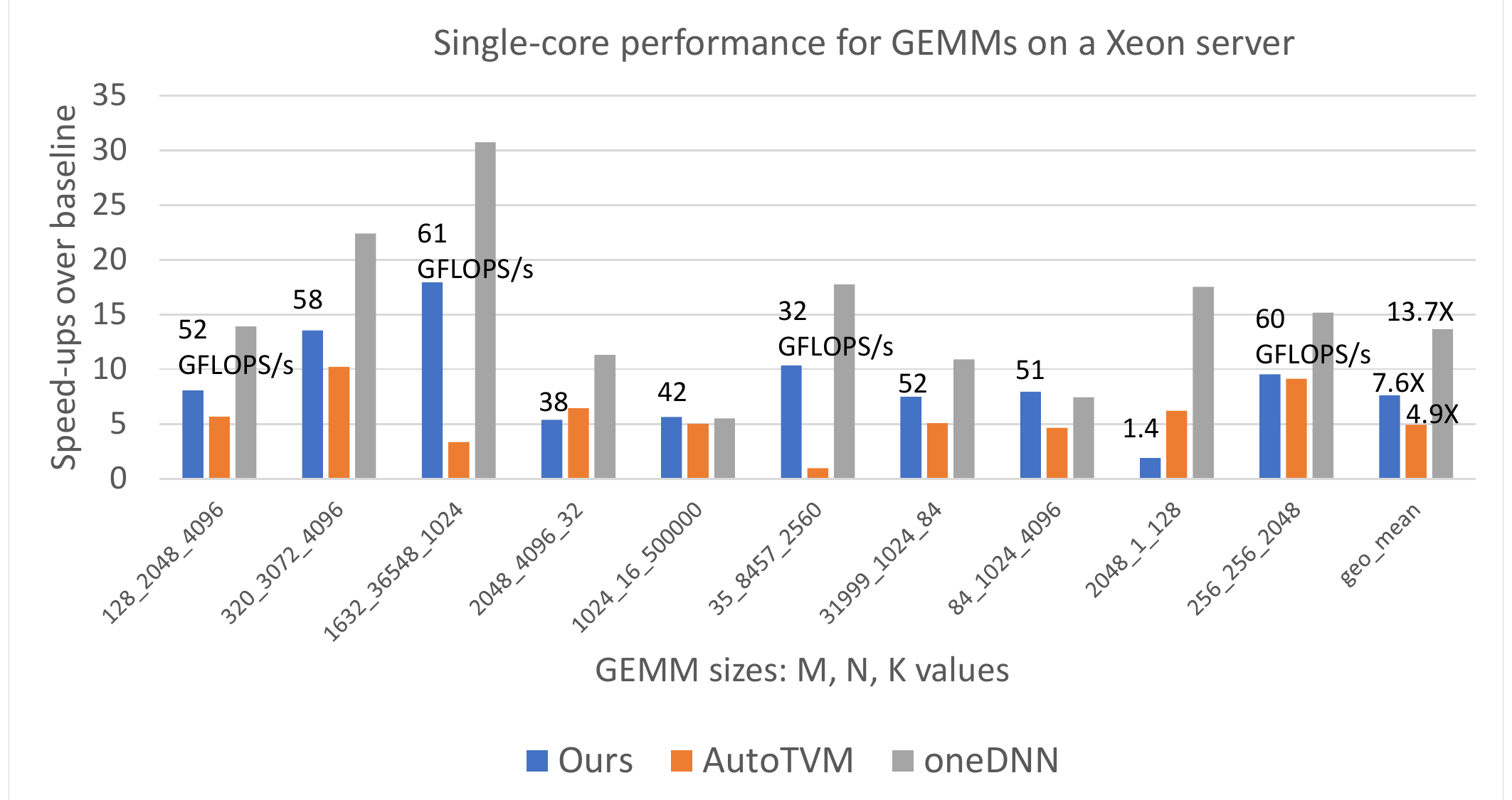}
\caption{The speed-ups achieved over the baseline by various systems. The absolute performances in GFLOPS/s obtained by our compiler toolchain for all problem sizes are shown.}
\label{fig:seqperf}
\end{figure*}

\subsection{Evaluation of High-level and Low-level optimizations together}
We next conduct experiments to evaluate how well our compiler toolchain performs
when we need to apply both high-level and low-level optimizations:
when matrix sizes are larger and therefore necessitate loop transformations including tiling to enhance data reuse in cache memories. 
Table \ref{tab:GEMMsizes} lists
the matrix sizes we perform the experiments with.
The matrix sizes are drawn from various deep learning applications.
We note that the matrix sizes are wide ranging
and therefore, test the versatility of our system in being able to come up with high
performance implementations for varied matrix sizes.

Two-level tiling is applied on the matrix multiplication code. Consequently,
there are six tile sizes that need to be selected.
We choose the best tile sizes using the high level optimization methodology
described in Section \ref{section:hlo}:
We first create a number of code variants by varying the tile sizes. 
The data reuse analyzer is run on each code variant and it outputs the working set sizes.
We execute the code variants on the target machine and measure their performance.
The working set sizes and the performance data are used together to 
train the DNN model for ranking of code variants.
While training the model, we use the training data corresponding to 70\% of the code variants.
Once the model is trained, for each GEMM size, we use the DNN model to select
the top 10\% best tile sizes from the space of candidate tile sizes.
For the top 10\% tile sizes thus selected, we apply low-level transformations (Section \ref{section:llo}) to vectorize the code.

\begin{figure*}
\centering
\includegraphics[scale=0.8]{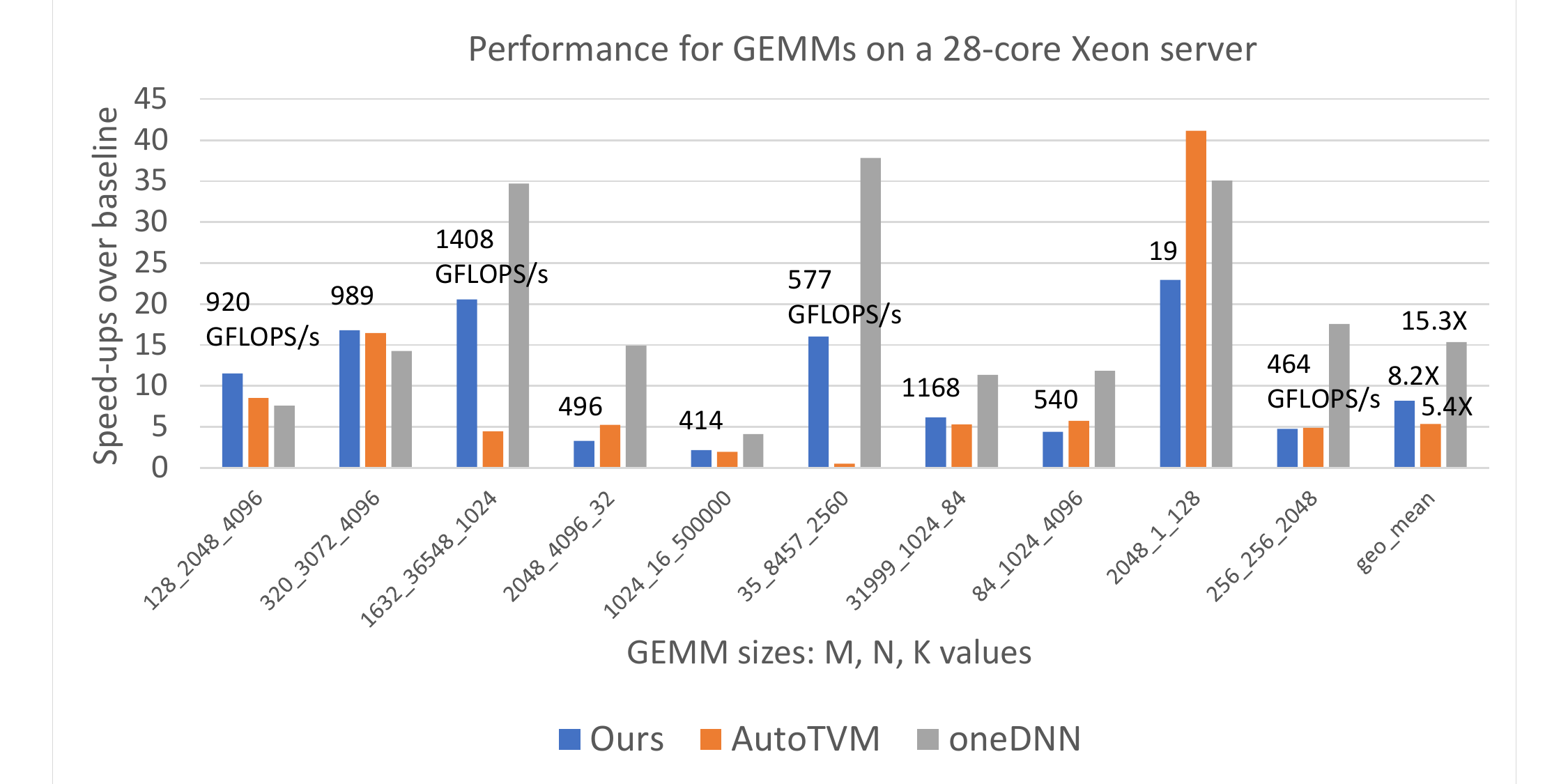}
\caption{The speed-ups achieved over the baseline by various systems. The absolute performances in GFLOPS/s obtained by our compiler toolchain for all problem sizes are shown.}
\label{fig:parallelperf}
\end{figure*}

Figure \ref{fig:seqperf} shows the speed-ups achieved over the baseline by our system, AutoTVM, and oneDNN library for sequential runs. We show the performance obtained by our compiler in absolute terms -- in terms of GFLOPS/s for each problem size.
When dealing with problem sizes that are odd numbers or when the optimal unroll factor for a loop
does not divide the corresponding tile size, then that leads to executing some parts of the computation
in scalar mode (and not vectorized). For example, the ``residue'' code  shown in Figure \ref{matmulunrolledAVX1}.  It leads to lower performance compared to when
everything is run in the vector mode. For example, for the problem sizes M = 31999,
N = 1024, and K = 84, the optimal unroll factors through the high-level and low-level
optimizations are determined to be 4, 32, and 1 for the three loops respectively. Because 4 does not divide 31999 exactly, it causes some part of the computation to be run
using scalar units. Our toolchain achieves
the highest performance of 60.7 GFLOPS/s for the problem size M = 1632, N = 36548, K = 1024. For this problem size, the highest speed-up relative to the baseline code  achieved is: 17.96X. The lowest performance of 1.4 GFLOPS/s is observed for M = 2048, N = 1, K = 128. The reason is, vectorization on the ``j'' loop as shown in Figure \ref{matmulunrolledAVX1} is not possible because the loop length of the ``j'' loop is 1.
If we were to vectorize on the ``i'' loop, that will induce non-contiguous data accesses for the A and C matrices and loading of data through vector-loads will not be possible.
That explains the low performance when N = 1.
We obtain on average (geometric mean) speed-up of 7.6X over the baseline.
The speed-ups achieved by the AutoTVM system and the oneDNN library are
4.9X and 13.7X respectively. In 8 out of 10 cases, our compiler outperforms AutoTVM.
In oneDNN implementations, data prefetch instructions are carefully inserted.
For example, low-level prefetch instructions such as \textsf{prefetcht0}
and \textsf{prefetcht2} which fetch data to all levels of cache and to L2 cache respectively are used extensively and that minimizes the number of instances when the processor has to wait for the arrival of data from memory.
Inter alia, such software data prefetching strategies employed in oneDNN explain its higher performance.

We measure the performance of parallel code when it is run on all 28 cores of the Xeon server. Figure \ref{fig:parallelperf} shows the speed-ups over the baseline for the three systems. The baseline code is parallelized as well. The average speed-ups over the baseline for our compiler, AutoTVM, and oneDNN are 8.2X, 5.4X, and 15.3X respectively.
For the first two problem sizes, namely M = 128, N = 2048, K = 4096, and M = 320, N = 3072, K = 4096, our compiler delivers higher performance even compared to oneDNN.
Our tool achieves the highest
performance of 1408 GFLOPS/s for the problem size M = 1632, N = 36548, K = 1024. Incidentally, the highest performance in sequential experiments is seen for the same problem size. For that problem size, it represents the parallel speed-up of 23.2X on a 28-core machine.
When the number of tiles of the parallelized loop is not a multiple of the number of cores (28), that leads to slight load imbalance among the cores. Consequently, we are hindered from achieving a perfect linear speed-up as the number of cores is increased.

\section{Related Work}
\label{sec:related}
In recent years, there has been a renewed interest in the application of Artificial Intelligence (A.I.) techniques for program optimization. 
The use of A.I. has been explored broadly for two purposes: 
1) for program representation in an embedding space \cite{ben2018neural,luan2019aroma,venkatakeerthy2020ir2vec,alon2019code2vec}, and 2) for performance
optimization \cite{chen2018tvm,chen2018learning,zheng2020flextensor,zheng2020ansor,franchetti2018spiral,park2013predictive}. In the former use case, once a program representation has been obtained then it has been used for tasks such as code comprehension, similar code search etc.
In the latter case when A.I. has been applied for program optimization, it has been used
to find optimal program transformations and optimal parameters for program transformations (such as loop unroll factors). Our work presented in the paper fits
the mold of the latter category where we use A.I. for performance enhancement.
Below, we describe several closely related works and detail how our present work improves upon and/or is different from prior works.

AutoTVM \cite{chen2018learning} uses machine learning approaches to derive
efficient execution schedules for tensor programs.
It explores the use of two distinct machine learning techniques:
In the first approach, from a given tensor program, domain specific features
such as memory access count, data reuse ratio etc are extracted.
Then, XGBoost, a form of decision tree based learners are trained to perform
relative ranking of schedules based on their performance.
In the second approach, the tensor program is encoded into an embedding vector
using the TreeGRU method \cite{tai2015improved}.
The relative performance prediction of the schedules is performed on the encoding thus obtained.

FlexTensor \cite{zheng2020flextensor} is a tensor computation optimization framework
for heterogeneous architectures. The hardware targets for which FlexTensor can create
high performance code include CPUs, GPUs, and FPGAs.
FlexTensor uses machine learning techniques to derive optimized execution schedules.
It uses two machine learning approaches: neural networks for performance
prediction, and reinforcement learning for navigating the space of possible schedules.
The schedule space is navigated using Q-learning \cite{watkins1992q} based reinforcement learning.
The Q-learning approach will guide the search -- along the directions of the search space
with the best performance.
The performance prediction for different schedules which is an input to the Q-learning algorithm
is performed using a feed-forward neural network.

Ansor \cite{zheng2020ansor} is a latest addition to the slew of TVM-based auto-tuning systems. 
Ansor considers a larger schedule space compared to the prior auto-tuning frameworks. 
Some of the innovations in Ansor include, 1) organization of the search space in a hierarchical manner 2) use of evolutionary search techniques 3) specification of schedules by recursive application of derivation rules 4) sampling of the search space -- by periodically running random schedules on the target hardware to better guide the search.


Park et al \cite{park2013predictive} use machine learning to select the best program transformation sequence among the available set of program transformation
recipes.
They run a given input program and obtain hardware counter values such as L1 cache misses.
Then, they also apply a sequence of program transformations and observe the achieved speed-up.
The hardware counters and the program transformations together are used as features for training a machine learning model
to predict the speed-ups.
Once the machine learning model is trained, it is used as follows for selecting the best program transformation sequence 
among a multitude of possibilities for a given program.
The unseen input program is run on the target hardware and the hardware counters are read.
The hardware counters along with a transformation sequence will be input to the trained machine learning model to predict
the speed-up that could be obtained for the transformation sequence.
The transformation sequence delivering the highest speed-up among various transformation sequences will be selected.

Various hand written implementations of basic linear algebra operations have been provided in various libraries like oneDNN \cite{intelmkldnn}, BLIS \cite{BLIS-10.1145/2755561}, OpenBLAS \cite{10.1109/ICPADS.2012.97}, GotoBLAS \cite{10.1145/1356052.1356053}.
ATLAS \cite{whaley1998automatically} is an autotuner where it generates various low-level C-implementations and finds the best performing one by executing the code on the target machine. POCA \cite{10.5555/3049832.3049846} generates LLVM-IR based vectorized GEMM micro-kernel where architecture independent optimizations can be  applied. AUGEM \cite{10.1145/2503210.2503219} is a template based code generator for DLA (Dense Linear algebra) operations. It replaces the common predefined C-code with the generated  assembly level code.
Kevin et al. \cite{10.1145/2086696.2086729} use assembly level features for analytical modeling of the SIMD code with Machine learning and find the best loop transformations for vectorization.
Monsifrot et al. \cite{10.5555/646053.677574} use a Machine learning approach to find the best unrolling factors.

Polyhedral compilation techniques have been developed for source-to-source code
transformation for better cache locality and parallelization.
The Pluto  compiler \cite{uday08pldi} derives an execution schedule
for the code that attempts to minimize data reuse distances.
The effect of the Pluto transformation will be that 
 iterations that use the same data will be executed close to each other in time
and therefore, the code will be cache friendly.
However, the performance obtained by polyhedral tools including Pluto's 
can be only slightly better than that of back-end compiler's such as Intel ICC's and far 
from other approaches such as AutoTVM's \cite{tavarageri2021polydl}.
In this paper, we show that our techniques outperform AutoTVM.
The reason for the inability of the purely polyhedral approaches is, they operate at a high level (source-to-source) and therefore,
do not perform low-level orchestration of vectorization, vector register allocation, detailed instruction scheduling (e.g., the kinds of low level optimizations we have described in Section \ref{section:llo}). The latter aspects are crucial to achieving high performance.

In our present work, we identified two distinct program optimizations that we need to concern ourselves with in order to obtain high performance on the target architectures.
Because the high-level and low-level optimizations are different, we developed different A.I. techniques for them. For high level optimizations we employed polyhedral compilation techniques to extract \emph{features} from loops and used a deep learning model to identify loop transformations that will yield high performance. For better vectorization, we combined an intrinsics based code generator with reinforcement learning (RL) to derive optimal parameters for the code generator. We have combined traditional compiler techniques with A.I. where appropriate. In particular, we have used A.I. where an accurate cost model is difficult to define and hence, we take advantage of A.I.'s unique ability to learn from performance data.

\section{Conclusion}
\label{sec:conclusion}
In this paper we presented CPU-focused compiler techniques for high-level and low-level program optimizations. The high-level optimizations enhance data locality of programs and the low-level transformations effectively vectorize code. A.I. techniques in conjunction with polyhedral compilation algorithms and target-specific code generator help us achieve high levels of performance. We demonstrated that matrix-multiplications which lie at the heart of deep learning can be effectively optimized using the developed compiler toolchain.

The presented approach will help re-target 
the toolchain to newer computer architectures
 seamlessly -- only a new target-specific code generator need be created, and the rest of
 the toolchain can be repurposed without modifications. 
Thus, the compiler framework described will enable realizing good performance out-of-the-box for new hardware architectures and new DL operators.

\balance
\bibliographystyle{ACM-Reference-Format}
\bibliography{paper}


\begin{thebibliography}{35}


\ifx \showCODEN    \undefined \def \showCODEN     #1{\unskip}     \fi
\ifx \showDOI      \undefined \def \showDOI       #1{#1}\fi
\ifx \showISBNx    \undefined \def \showISBNx     #1{\unskip}     \fi
\ifx \showISBNxiii \undefined \def \showISBNxiii  #1{\unskip}     \fi
\ifx \showISSN     \undefined \def \showISSN      #1{\unskip}     \fi
\ifx \showLCCN     \undefined \def \showLCCN      #1{\unskip}     \fi
\ifx \shownote     \undefined \def \shownote      #1{#1}          \fi
\ifx \showarticletitle \undefined \def \showarticletitle #1{#1}   \fi
\ifx \showURL      \undefined \def \showURL       {\relax}        \fi
\providecommand\bibfield[2]{#2}
\providecommand\bibinfo[2]{#2}
\providecommand\natexlab[1]{#1}
\providecommand\showeprint[2][]{arXiv:#2}

\bibitem[\protect\citeauthoryear{??}{gem}{2015}]%
        {gemmdl}
 \bibinfo{year}{2015}\natexlab{}.
\newblock \bibinfo{booktitle}{\emph{Why GEMM is at the heart of deep
  learning}}.
\newblock
\urldef\tempurl%
\url{https://petewarden.com/2015/04/20/why-gemm-is-at-the-heart-of-deep-learning/}
\showURL{%
\tempurl}


\bibitem[\protect\citeauthoryear{??}{aut}{2020}]%
        {autotvmgemmcpu}
 \bibinfo{year}{2020}\natexlab{}.
\newblock \bibinfo{booktitle}{\emph{How to optimize GEMM on CPU}}.
\newblock
\urldef\tempurl%
\url{https://tvm.apache.org/docs/tutorials/optimize/opt\_gemm.html}
\showURL{%
\tempurl}


\bibitem[\protect\citeauthoryear{??}{int}{2020}]%
        {intelmkldnn}
 \bibinfo{year}{2020}\natexlab{}.
\newblock \bibinfo{title}{oneAPI Deep Neural Network Library (oneDNN)}.
\newblock
\newblock
\urldef\tempurl%
\url{https://github.com/oneapi-src/oneDNN}
\showURL{%
\tempurl}


\bibitem[\protect\citeauthoryear{Alon, Zilberstein, Levy, and Yahav}{Alon
  et~al\mbox{.}}{2019}]%
        {alon2019code2vec}
\bibfield{author}{\bibinfo{person}{Uri Alon}, \bibinfo{person}{Meital
  Zilberstein}, \bibinfo{person}{Omer Levy}, {and} \bibinfo{person}{Eran
  Yahav}.} \bibinfo{year}{2019}\natexlab{}.
\newblock \showarticletitle{code2vec: Learning distributed representations of
  code}.
\newblock \bibinfo{journal}{\emph{Proceedings of the ACM on Programming
  Languages}} \bibinfo{volume}{3}, \bibinfo{number}{POPL}
  (\bibinfo{year}{2019}), \bibinfo{pages}{1--29}.
\newblock


\bibitem[\protect\citeauthoryear{Batra, Jacobson, Madhav, Queirolo, and
  Santhanam}{Batra et~al\mbox{.}}{2018}]%
        {mckinseystudyinferencehardware}
\bibfield{author}{\bibinfo{person}{Gaurav Batra}, \bibinfo{person}{Zach
  Jacobson}, \bibinfo{person}{Siddarth Madhav}, \bibinfo{person}{Andrea
  Queirolo}, {and} \bibinfo{person}{Nick Santhanam}.}
  \bibinfo{year}{2018}\natexlab{}.
\newblock \bibinfo{booktitle}{\emph{Artificial-intelligence hardware: New
  opportunities for semiconductor companies}}.
\newblock
\urldef\tempurl%
\url{https://www.mckinsey.com/industries/semiconductors/our-insights}
\showURL{%
\tempurl}


\bibitem[\protect\citeauthoryear{Ben-Nun, Jakobovits, and Hoefler}{Ben-Nun
  et~al\mbox{.}}{2018}]%
        {ben2018neural}
\bibfield{author}{\bibinfo{person}{Tal Ben-Nun},
  \bibinfo{person}{Alice~Shoshana Jakobovits}, {and} \bibinfo{person}{Torsten
  Hoefler}.} \bibinfo{year}{2018}\natexlab{}.
\newblock \showarticletitle{Neural code comprehension: A learnable
  representation of code semantics}.
\newblock \bibinfo{journal}{\emph{Advances in Neural Information Processing
  Systems}}  \bibinfo{volume}{31} (\bibinfo{year}{2018}),
  \bibinfo{pages}{3585--3597}.
\newblock


\bibitem[\protect\citeauthoryear{Bondhugula, Hartono, Ramanujam, and
  Sadayappan}{Bondhugula et~al\mbox{.}}{2008}]%
        {uday08pldi}
\bibfield{author}{\bibinfo{person}{Uday Bondhugula}, \bibinfo{person}{Albert
  Hartono}, \bibinfo{person}{J. Ramanujam}, {and} \bibinfo{person}{P.
  Sadayappan}.} \bibinfo{year}{2008}\natexlab{}.
\newblock \showarticletitle{A Practical Automatic Polyhedral Program
  Optimization System}. In \bibinfo{booktitle}{\emph{ACM SIGPLAN Conference on
  Programming Language Design and Implementation (PLDI)}}.
\newblock


\bibitem[\protect\citeauthoryear{Chen, Moreau, Jiang, Zheng, Yan, Shen, Cowan,
  Wang, Hu, Ceze, et~al\mbox{.}}{Chen et~al\mbox{.}}{2018a}]%
        {chen2018tvm}
\bibfield{author}{\bibinfo{person}{Tianqi Chen}, \bibinfo{person}{Thierry
  Moreau}, \bibinfo{person}{Ziheng Jiang}, \bibinfo{person}{Lianmin Zheng},
  \bibinfo{person}{Eddie Yan}, \bibinfo{person}{Haichen Shen},
  \bibinfo{person}{Meghan Cowan}, \bibinfo{person}{Leyuan Wang},
  \bibinfo{person}{Yuwei Hu}, \bibinfo{person}{Luis Ceze}, {et~al\mbox{.}}}
  \bibinfo{year}{2018}\natexlab{a}.
\newblock \showarticletitle{$\{$TVM$\}$: An automated end-to-end optimizing
  compiler for deep learning}. In \bibinfo{booktitle}{\emph{13th $\{$USENIX$\}$
  Symposium on Operating Systems Design and Implementation ($\{$OSDI$\}$ 18)}}.
  \bibinfo{pages}{578--594}.
\newblock


\bibitem[\protect\citeauthoryear{Chen, Zheng, Yan, Jiang, Moreau, Ceze,
  Guestrin, and Krishnamurthy}{Chen et~al\mbox{.}}{2018b}]%
        {chen2018learning}
\bibfield{author}{\bibinfo{person}{Tianqi Chen}, \bibinfo{person}{Lianmin
  Zheng}, \bibinfo{person}{Eddie Yan}, \bibinfo{person}{Ziheng Jiang},
  \bibinfo{person}{Thierry Moreau}, \bibinfo{person}{Luis Ceze},
  \bibinfo{person}{Carlos Guestrin}, {and} \bibinfo{person}{Arvind
  Krishnamurthy}.} \bibinfo{year}{2018}\natexlab{b}.
\newblock \showarticletitle{Learning to optimize tensor programs}. In
  \bibinfo{booktitle}{\emph{Advances in Neural Information Processing
  Systems}}. \bibinfo{pages}{3389--3400}.
\newblock


\bibitem[\protect\citeauthoryear{Devlin, Chang, Lee, and Toutanova}{Devlin
  et~al\mbox{.}}{2018}]%
        {devlin2018bert}
\bibfield{author}{\bibinfo{person}{Jacob Devlin}, \bibinfo{person}{Ming-Wei
  Chang}, \bibinfo{person}{Kenton Lee}, {and} \bibinfo{person}{Kristina
  Toutanova}.} \bibinfo{year}{2018}\natexlab{}.
\newblock \showarticletitle{Bert: Pre-training of deep bidirectional
  transformers for language understanding}.
\newblock \bibinfo{journal}{\emph{arXiv preprint arXiv:1810.04805}}
  (\bibinfo{year}{2018}).
\newblock


\bibitem[\protect\citeauthoryear{Feautrier}{Feautrier}{1996}]%
        {feautrier1996automatic}
\bibfield{author}{\bibinfo{person}{Paul Feautrier}.}
  \bibinfo{year}{1996}\natexlab{}.
\newblock \showarticletitle{Automatic parallelization in the polytope model}.
\newblock In \bibinfo{booktitle}{\emph{The Data Parallel Programming Model}}.
  \bibinfo{publisher}{Springer}, \bibinfo{pages}{79--103}.
\newblock


\bibitem[\protect\citeauthoryear{Franchetti, Low, Popovici, Veras, Spampinato,
  Johnson, P{\"u}schel, Hoe, and Moura}{Franchetti et~al\mbox{.}}{2018}]%
        {franchetti2018spiral}
\bibfield{author}{\bibinfo{person}{Franz Franchetti}, \bibinfo{person}{Tze~Meng
  Low}, \bibinfo{person}{Doru~Thom Popovici}, \bibinfo{person}{Richard~M
  Veras}, \bibinfo{person}{Daniele~G Spampinato}, \bibinfo{person}{Jeremy~R
  Johnson}, \bibinfo{person}{Markus P{\"u}schel}, \bibinfo{person}{James~C
  Hoe}, {and} \bibinfo{person}{Jos{\'e}~MF Moura}.}
  \bibinfo{year}{2018}\natexlab{}.
\newblock \showarticletitle{SPIRAL: Extreme performance portability}.
\newblock \bibinfo{journal}{\emph{Proc. IEEE}} \bibinfo{volume}{106},
  \bibinfo{number}{11} (\bibinfo{year}{2018}), \bibinfo{pages}{1935--1968}.
\newblock


\bibitem[\protect\citeauthoryear{Goto and Geijn}{Goto and Geijn}{2008}]%
        {10.1145/1356052.1356053}
\bibfield{author}{\bibinfo{person}{Kazushige Goto} {and}
  \bibinfo{person}{Robert A. van~de Geijn}.} \bibinfo{year}{2008}\natexlab{}.
\newblock \showarticletitle{Anatomy of High-Performance Matrix Multiplication}.
\newblock \bibinfo{journal}{\emph{ACM Trans. Math. Softw.}}
  \bibinfo{volume}{34}, \bibinfo{number}{3}, Article \bibinfo{articleno}{12}
  (\bibinfo{date}{May} \bibinfo{year}{2008}), \bibinfo{numpages}{25}~pages.
\newblock
\showISSN{0098-3500}
\urldef\tempurl%
\url{https://doi.org/10.1145/1356052.1356053}
\showDOI{\tempurl}


\bibitem[\protect\citeauthoryear{He, Zhang, Ren, and Sun}{He
  et~al\mbox{.}}{2016}]%
        {he2016deep}
\bibfield{author}{\bibinfo{person}{Kaiming He}, \bibinfo{person}{Xiangyu
  Zhang}, \bibinfo{person}{Shaoqing Ren}, {and} \bibinfo{person}{Jian Sun}.}
  \bibinfo{year}{2016}\natexlab{}.
\newblock \showarticletitle{Deep residual learning for image recognition}. In
  \bibinfo{booktitle}{\emph{Proceedings of the IEEE conference on computer
  vision and pattern recognition}}. \bibinfo{pages}{770--778}.
\newblock


\bibitem[\protect\citeauthoryear{Hinton, Deng, Yu, Dahl, Mohamed, Jaitly,
  Senior, Vanhoucke, Nguyen, Kingsbury, et~al\mbox{.}}{Hinton
  et~al\mbox{.}}{2012}]%
        {hinton2012deep}
\bibfield{author}{\bibinfo{person}{Geoffrey Hinton}, \bibinfo{person}{Li Deng},
  \bibinfo{person}{Dong Yu}, \bibinfo{person}{George Dahl},
  \bibinfo{person}{Abdel-rahman Mohamed}, \bibinfo{person}{Navdeep Jaitly},
  \bibinfo{person}{Andrew Senior}, \bibinfo{person}{Vincent Vanhoucke},
  \bibinfo{person}{Patrick Nguyen}, \bibinfo{person}{Brian Kingsbury},
  {et~al\mbox{.}}} \bibinfo{year}{2012}\natexlab{}.
\newblock \showarticletitle{Deep neural networks for acoustic modeling in
  speech recognition}.
\newblock \bibinfo{journal}{\emph{IEEE Signal processing magazine}}
  \bibinfo{volume}{29} (\bibinfo{year}{2012}).
\newblock


\bibitem[\protect\citeauthoryear{Jouppi, Young, Patil, Patterson, Agrawal,
  Bajwa, Bates, Bhatia, Boden, Borchers, et~al\mbox{.}}{Jouppi
  et~al\mbox{.}}{2017}]%
        {jouppi2017datacenter}
\bibfield{author}{\bibinfo{person}{Norman~P Jouppi}, \bibinfo{person}{Cliff
  Young}, \bibinfo{person}{Nishant Patil}, \bibinfo{person}{David Patterson},
  \bibinfo{person}{Gaurav Agrawal}, \bibinfo{person}{Raminder Bajwa},
  \bibinfo{person}{Sarah Bates}, \bibinfo{person}{Suresh Bhatia},
  \bibinfo{person}{Nan Boden}, \bibinfo{person}{Al Borchers}, {et~al\mbox{.}}}
  \bibinfo{year}{2017}\natexlab{}.
\newblock \showarticletitle{In-datacenter performance analysis of a tensor
  processing unit}. In \bibinfo{booktitle}{\emph{2017 ACM/IEEE 44th Annual
  International Symposium on Computer Architecture (ISCA)}}. IEEE,
  \bibinfo{pages}{1--12}.
\newblock


\bibitem[\protect\citeauthoryear{Krizhevsky, Sutskever, and Hinton}{Krizhevsky
  et~al\mbox{.}}{2012}]%
        {krizhevsky2012imagenet}
\bibfield{author}{\bibinfo{person}{Alex Krizhevsky}, \bibinfo{person}{Ilya
  Sutskever}, {and} \bibinfo{person}{Geoffrey~E Hinton}.}
  \bibinfo{year}{2012}\natexlab{}.
\newblock \showarticletitle{Imagenet classification with deep convolutional
  neural networks}. In \bibinfo{booktitle}{\emph{Advances in neural information
  processing systems}}. \bibinfo{pages}{1097--1105}.
\newblock


\bibitem[\protect\citeauthoryear{Luan, Yang, Barnaby, Sen, and Chandra}{Luan
  et~al\mbox{.}}{2019}]%
        {luan2019aroma}
\bibfield{author}{\bibinfo{person}{Sifei Luan}, \bibinfo{person}{Di Yang},
  \bibinfo{person}{Celeste Barnaby}, \bibinfo{person}{Koushik Sen}, {and}
  \bibinfo{person}{Satish Chandra}.} \bibinfo{year}{2019}\natexlab{}.
\newblock \showarticletitle{Aroma: Code recommendation via structural code
  search}.
\newblock \bibinfo{journal}{\emph{Proceedings of the ACM on Programming
  Languages}} \bibinfo{volume}{3}, \bibinfo{number}{OOPSLA}
  (\bibinfo{year}{2019}), \bibinfo{pages}{1--28}.
\newblock


\bibitem[\protect\citeauthoryear{Monsifrot, Bodin, and Quiniou}{Monsifrot
  et~al\mbox{.}}{2002}]%
        {10.5555/646053.677574}
\bibfield{author}{\bibinfo{person}{Antoine Monsifrot},
  \bibinfo{person}{Fran\c{c}ois Bodin}, {and} \bibinfo{person}{Rene Quiniou}.}
  \bibinfo{year}{2002}\natexlab{}.
\newblock \showarticletitle{A Machine Learning Approach to Automatic Production
  of Compiler Heuristics}. In \bibinfo{booktitle}{\emph{Proceedings of the 10th
  International Conference on Artificial Intelligence: Methodology, Systems,
  and Applications}} \emph{(\bibinfo{series}{AIMSA '02})}.
  \bibinfo{publisher}{Springer-Verlag}, \bibinfo{address}{Berlin, Heidelberg},
  \bibinfo{pages}{41–50}.
\newblock
\showISBNx{3540441271}


\bibitem[\protect\citeauthoryear{Park, Cavazos, Pouchet, Bastoul, Cohen, and
  Sadayappan}{Park et~al\mbox{.}}{2013}]%
        {park2013predictive}
\bibfield{author}{\bibinfo{person}{Eunjung Park}, \bibinfo{person}{John
  Cavazos}, \bibinfo{person}{Louis-No{\"e}l Pouchet},
  \bibinfo{person}{C{\'e}dric Bastoul}, \bibinfo{person}{Albert Cohen}, {and}
  \bibinfo{person}{P Sadayappan}.} \bibinfo{year}{2013}\natexlab{}.
\newblock \showarticletitle{Predictive modeling in a polyhedral optimization
  space}.
\newblock \bibinfo{journal}{\emph{International journal of parallel
  programming}} \bibinfo{volume}{41}, \bibinfo{number}{5}
  (\bibinfo{year}{2013}), \bibinfo{pages}{704--750}.
\newblock


\bibitem[\protect\citeauthoryear{Qin, Samajdar, Kwon, Nadella, Srinivasan, Das,
  Kaul, and Krishna}{Qin et~al\mbox{.}}{2020}]%
        {qin2020sigma}
\bibfield{author}{\bibinfo{person}{Eric Qin}, \bibinfo{person}{Ananda
  Samajdar}, \bibinfo{person}{Hyoukjun Kwon}, \bibinfo{person}{Vineet Nadella},
  \bibinfo{person}{Sudarshan Srinivasan}, \bibinfo{person}{Dipankar Das},
  \bibinfo{person}{Bharat Kaul}, {and} \bibinfo{person}{Tushar Krishna}.}
  \bibinfo{year}{2020}\natexlab{}.
\newblock \showarticletitle{Sigma: A sparse and irregular gemm accelerator with
  flexible interconnects for dnn training}. In \bibinfo{booktitle}{\emph{2020
  IEEE International Symposium on High Performance Computer Architecture
  (HPCA)}}. IEEE, \bibinfo{pages}{58--70}.
\newblock


\bibitem[\protect\citeauthoryear{Stock, Pouchet, and Sadayappan}{Stock
  et~al\mbox{.}}{2012}]%
        {10.1145/2086696.2086729}
\bibfield{author}{\bibinfo{person}{Kevin Stock},
  \bibinfo{person}{Louis-No\"{e}l Pouchet}, {and} \bibinfo{person}{P.
  Sadayappan}.} \bibinfo{year}{2012}\natexlab{}.
\newblock \showarticletitle{Using Machine Learning to Improve Automatic
  Vectorization}.
\newblock \bibinfo{journal}{\emph{ACM Trans. Archit. Code Optim.}}
  \bibinfo{volume}{8}, \bibinfo{number}{4}, Article \bibinfo{articleno}{50}
  (\bibinfo{date}{Jan.} \bibinfo{year}{2012}), \bibinfo{numpages}{23}~pages.
\newblock
\showISSN{1544-3566}
\urldef\tempurl%
\url{https://doi.org/10.1145/2086696.2086729}
\showDOI{\tempurl}


\bibitem[\protect\citeauthoryear{Su, Liao, and Xue}{Su et~al\mbox{.}}{2017}]%
        {10.5555/3049832.3049846}
\bibfield{author}{\bibinfo{person}{Xing Su}, \bibinfo{person}{Xiangke Liao},
  {and} \bibinfo{person}{Jingling Xue}.} \bibinfo{year}{2017}\natexlab{}.
\newblock \showarticletitle{Automatic Generation of Fast BLAS3-GEMM: A Portable
  Compiler Approach}. In \bibinfo{booktitle}{\emph{Proceedings of the 2017
  International Symposium on Code Generation and Optimization}} (Austin, USA)
  \emph{(\bibinfo{series}{CGO '17})}. \bibinfo{publisher}{IEEE Press},
  \bibinfo{pages}{122–133}.
\newblock
\showISBNx{9781509049318}


\bibitem[\protect\citeauthoryear{Tai, Socher, and Manning}{Tai
  et~al\mbox{.}}{2015}]%
        {tai2015improved}
\bibfield{author}{\bibinfo{person}{Kai~Sheng Tai}, \bibinfo{person}{Richard
  Socher}, {and} \bibinfo{person}{Christopher~D Manning}.}
  \bibinfo{year}{2015}\natexlab{}.
\newblock \showarticletitle{Improved semantic representations from
  tree-structured long short-term memory networks}.
\newblock \bibinfo{journal}{\emph{arXiv preprint arXiv:1503.00075}}
  (\bibinfo{year}{2015}).
\newblock


\bibitem[\protect\citeauthoryear{Tavarageri, Heinecke, Avancha, Kaul, Goyal,
  and Upadrasta}{Tavarageri et~al\mbox{.}}{2021}]%
        {tavarageri2021polydl}
\bibfield{author}{\bibinfo{person}{Sanket Tavarageri},
  \bibinfo{person}{Alexander Heinecke}, \bibinfo{person}{Sasikanth Avancha},
  \bibinfo{person}{Bharat Kaul}, \bibinfo{person}{Gagandeep Goyal}, {and}
  \bibinfo{person}{Ramakrishna Upadrasta}.} \bibinfo{year}{2021}\natexlab{}.
\newblock \showarticletitle{PolyDL: Polyhedral Optimizations for Creation of
  High-performance DL Primitives}.
\newblock \bibinfo{journal}{\emph{ACM Transactions on Architecture and Code
  Optimization (TACO)}} \bibinfo{volume}{18}, \bibinfo{number}{1}
  (\bibinfo{year}{2021}), \bibinfo{pages}{1--27}.
\newblock


\bibitem[\protect\citeauthoryear{VenkataKeerthy, Aggarwal, Jain, Desarkar,
  Upadrasta, and Srikant}{VenkataKeerthy et~al\mbox{.}}{2020}]%
        {venkatakeerthy2020ir2vec}
\bibfield{author}{\bibinfo{person}{S VenkataKeerthy}, \bibinfo{person}{Rohit
  Aggarwal}, \bibinfo{person}{Shalini Jain}, \bibinfo{person}{Maunendra~Sankar
  Desarkar}, \bibinfo{person}{Ramakrishna Upadrasta}, {and} \bibinfo{person}{YN
  Srikant}.} \bibinfo{year}{2020}\natexlab{}.
\newblock \showarticletitle{IR2Vec: LLVM IR Based Scalable Program Embeddings}.
\newblock \bibinfo{journal}{\emph{ACM Transactions on Architecture and Code
  Optimization (TACO)}} \bibinfo{volume}{17}, \bibinfo{number}{4}
  (\bibinfo{year}{2020}), \bibinfo{pages}{1--27}.
\newblock


\bibitem[\protect\citeauthoryear{Verdoolaege}{Verdoolaege}{2010}]%
        {verdoolaege2010isl}
\bibfield{author}{\bibinfo{person}{Sven Verdoolaege}.}
  \bibinfo{year}{2010}\natexlab{}.
\newblock \showarticletitle{isl: An integer set library for the polyhedral
  model}. In \bibinfo{booktitle}{\emph{International Congress on Mathematical
  Software}}. Springer, \bibinfo{pages}{299--302}.
\newblock


\bibitem[\protect\citeauthoryear{Wang, Zhang, Zhang, and Yi}{Wang
  et~al\mbox{.}}{2013}]%
        {10.1145/2503210.2503219}
\bibfield{author}{\bibinfo{person}{Qian Wang}, \bibinfo{person}{Xianyi Zhang},
  \bibinfo{person}{Yunquan Zhang}, {and} \bibinfo{person}{Qing Yi}.}
  \bibinfo{year}{2013}\natexlab{}.
\newblock \showarticletitle{AUGEM: Automatically Generate High Performance
  Dense Linear Algebra Kernels on X86 CPUs}. In
  \bibinfo{booktitle}{\emph{Proceedings of the International Conference on High
  Performance Computing, Networking, Storage and Analysis}} (Denver, Colorado)
  \emph{(\bibinfo{series}{SC '13})}. \bibinfo{publisher}{Association for
  Computing Machinery}, \bibinfo{address}{New York, NY, USA}, Article
  \bibinfo{articleno}{25}, \bibinfo{numpages}{12}~pages.
\newblock
\showISBNx{9781450323789}
\urldef\tempurl%
\url{https://doi.org/10.1145/2503210.2503219}
\showDOI{\tempurl}


\bibitem[\protect\citeauthoryear{Watkins and Dayan}{Watkins and Dayan}{1992}]%
        {watkins1992q}
\bibfield{author}{\bibinfo{person}{Christopher~JCH Watkins} {and}
  \bibinfo{person}{Peter Dayan}.} \bibinfo{year}{1992}\natexlab{}.
\newblock \showarticletitle{Q-learning}.
\newblock \bibinfo{journal}{\emph{Machine learning}} \bibinfo{volume}{8},
  \bibinfo{number}{3-4} (\bibinfo{year}{1992}), \bibinfo{pages}{279--292}.
\newblock


\bibitem[\protect\citeauthoryear{Whaley and Dongarra}{Whaley and
  Dongarra}{1998}]%
        {whaley1998automatically}
\bibfield{author}{\bibinfo{person}{R~Clinton Whaley} {and}
  \bibinfo{person}{Jack~J Dongarra}.} \bibinfo{year}{1998}\natexlab{}.
\newblock \showarticletitle{Automatically tuned linear algebra software}. In
  \bibinfo{booktitle}{\emph{SC'98: Proceedings of the 1998 ACM/IEEE conference
  on Supercomputing}}. IEEE, \bibinfo{pages}{38--38}.
\newblock


\bibitem[\protect\citeauthoryear{Wu, Schuster, Chen, Le, Norouzi, Macherey,
  Krikun, Cao, Gao, Macherey, et~al\mbox{.}}{Wu et~al\mbox{.}}{2016}]%
        {wu2016google}
\bibfield{author}{\bibinfo{person}{Yonghui Wu}, \bibinfo{person}{Mike
  Schuster}, \bibinfo{person}{Zhifeng Chen}, \bibinfo{person}{Quoc~V Le},
  \bibinfo{person}{Mohammad Norouzi}, \bibinfo{person}{Wolfgang Macherey},
  \bibinfo{person}{Maxim Krikun}, \bibinfo{person}{Yuan Cao},
  \bibinfo{person}{Qin Gao}, \bibinfo{person}{Klaus Macherey}, {et~al\mbox{.}}}
  \bibinfo{year}{2016}\natexlab{}.
\newblock \showarticletitle{Google's neural machine translation system:
  Bridging the gap between human and machine translation}.
\newblock \bibinfo{journal}{\emph{arXiv preprint arXiv:1609.08144}}
  (\bibinfo{year}{2016}).
\newblock


\bibitem[\protect\citeauthoryear{Xianyi, Qian, and Yunquan}{Xianyi
  et~al\mbox{.}}{2012}]%
        {10.1109/ICPADS.2012.97}
\bibfield{author}{\bibinfo{person}{Zhang Xianyi}, \bibinfo{person}{Wang Qian},
  {and} \bibinfo{person}{Zhang Yunquan}.} \bibinfo{year}{2012}\natexlab{}.
\newblock \showarticletitle{Model-Driven Level 3 BLAS Performance Optimization
  on Loongson 3A Processor}. In \bibinfo{booktitle}{\emph{Proceedings of the
  2012 IEEE 18th International Conference on Parallel and Distributed Systems}}
  \emph{(\bibinfo{series}{ICPADS '12})}. \bibinfo{publisher}{IEEE Computer
  Society}, \bibinfo{address}{USA}, \bibinfo{pages}{684–691}.
\newblock
\showISBNx{9780769549033}
\urldef\tempurl%
\url{https://doi.org/10.1109/ICPADS.2012.97}
\showDOI{\tempurl}


\bibitem[\protect\citeauthoryear{Zee, Smith, Marker, Low, Geijn, Igual,
  Smelyanskiy, Zhang, Kistler, Austel, Gunnels, and Killough}{Zee
  et~al\mbox{.}}{2016}]%
        {BLIS-10.1145/2755561}
\bibfield{author}{\bibinfo{person}{Field G.~Van Zee}, \bibinfo{person}{Tyler~M.
  Smith}, \bibinfo{person}{Bryan Marker}, \bibinfo{person}{Tze~Meng Low},
  \bibinfo{person}{Robert A. Van~De Geijn}, \bibinfo{person}{Francisco~D.
  Igual}, \bibinfo{person}{Mikhail Smelyanskiy}, \bibinfo{person}{Xianyi
  Zhang}, \bibinfo{person}{Michael Kistler}, \bibinfo{person}{Vernon Austel},
  \bibinfo{person}{John~A. Gunnels}, {and} \bibinfo{person}{Lee Killough}.}
  \bibinfo{year}{2016}\natexlab{}.
\newblock \showarticletitle{The BLIS Framework: Experiments in Portability}.
\newblock \bibinfo{journal}{\emph{ACM Trans. Math. Softw.}}
  \bibinfo{volume}{42}, \bibinfo{number}{2}, Article \bibinfo{articleno}{12}
  (\bibinfo{date}{June} \bibinfo{year}{2016}), \bibinfo{numpages}{19}~pages.
\newblock
\showISSN{0098-3500}
\urldef\tempurl%
\url{https://doi.org/10.1145/2755561}
\showDOI{\tempurl}


\bibitem[\protect\citeauthoryear{Zheng, Jia, Sun, Wu, Yu, Haj-Ali, Wang, Yang,
  Zhuo, Sen, et~al\mbox{.}}{Zheng et~al\mbox{.}}{2020a}]%
        {zheng2020ansor}
\bibfield{author}{\bibinfo{person}{Lianmin Zheng}, \bibinfo{person}{Chengfan
  Jia}, \bibinfo{person}{Minmin Sun}, \bibinfo{person}{Zhao Wu},
  \bibinfo{person}{Cody~Hao Yu}, \bibinfo{person}{Ameer Haj-Ali},
  \bibinfo{person}{Yida Wang}, \bibinfo{person}{Jun Yang},
  \bibinfo{person}{Danyang Zhuo}, \bibinfo{person}{Koushik Sen},
  {et~al\mbox{.}}} \bibinfo{year}{2020}\natexlab{a}.
\newblock \showarticletitle{Ansor: Generating high-performance tensor programs
  for deep learning}. In \bibinfo{booktitle}{\emph{14th $\{$USENIX$\}$
  Symposium on Operating Systems Design and Implementation ($\{$OSDI$\}$ 20)}}.
  \bibinfo{pages}{863--879}.
\newblock


\bibitem[\protect\citeauthoryear{Zheng, Liang, Wang, Chen, and Sheng}{Zheng
  et~al\mbox{.}}{2020b}]%
        {zheng2020flextensor}
\bibfield{author}{\bibinfo{person}{Size Zheng}, \bibinfo{person}{Yun Liang},
  \bibinfo{person}{Shuo Wang}, \bibinfo{person}{Renze Chen}, {and}
  \bibinfo{person}{Kaiwen Sheng}.} \bibinfo{year}{2020}\natexlab{b}.
\newblock \showarticletitle{FlexTensor: An Automatic Schedule Exploration and
  Optimization Framework for Tensor Computation on Heterogeneous System}. In
  \bibinfo{booktitle}{\emph{Proceedings of the Twenty-Fifth International
  Conference on Architectural Support for Programming Languages and Operating
  Systems}}. \bibinfo{pages}{859--873}.
\newblock


\end{thebibliography}


\end{document}